\documentclass[twocolumn,showpacs,amsmath,amssymb]{revtex4}

\usepackage{epsfig}
              
\usepackage{dcolumn}

\usepackage{bm}

\usepackage{amsmath}

\usepackage{amsfonts}

\usepackage{amssymb}

\usepackage{graphicx}%

\newcommand{\beq}{\begin{eqnarray}}

\newcommand{\eeq}{\end{eqnarray}}

\begin{document}

\title{Heat production and energy balance in nanoengines driven
  by time-dependent fields.}

\author{Liliana Arrachea$^{(1),(2)}$, Michael Moskalets$^{(3)}$ and Luis Martin-Moreno$^{(1),(2)}$}
\affiliation{$^{(1)}$ Departamento de F\'{\i}sica de la Materia Condensada,
Universidad de Zaragoza, Pedro Cerbuna 12 (50009) Zaragoza, Spain,\\
$^{(2)}$Instituto de Biocomputaci\'on y F\'{\i}sica de Sistemas
  Complejos, Universidad de Zaragoza,
Corona de Arag\'on 42, (50009) Zaragoza, Spain, \\
$^{(3)}$ Department of Metal and Semiconductor Physics, National Technical
University, ``Kharkov Polytechnical Institute'', (61002) Kharkov, Ukraine.}


\begin{abstract}
We present a formalism to study the heat transport  and the power developed
 by the local driving fields on a quantum system coupled
 to macroscopic reservoirs.
We show  that, quite generally, two important mechanisms can take place:
 (i) directed heat transport between reservoirs induced by the ac potentials
and (ii)
at slow driving, two oscillating out
 of phase forces  perform work against each other,
while the energy dissipated into  the reservoirs is negligible.
\end{abstract}

\maketitle

The understanding of the heat transport at the microscopic realm has
attracted the attention of theoreticians for several years now.
Several studies investigate this issue
in the framework of one-dimensional lattice models of interacting
classical oscillators. \cite{fer,cam,gen,giar,pere} Nowadays,
the technological trend towards the fabrication of nanosize electronic
devices, is boosting the theoretical
interest in
quantum transport
in a variety of setups and
materials. Recently,
 there have been efforts to address
the related problems of energy transport and heat dissipation in these
small-size
systems. \cite{AEGS01,mobu,wan,linkeheat,nitzan,bee,rey}

Electronic quantum transport through mesoscopic systems has been traditionally
analyzed as a response to dc-voltages.
 There are, however, alternative
possibilities to induce net transport by using
time-dependent fields as the generating source. Interesting
examples of this kind have been recently realized experimentally.  \cite{pumpex,saw}
An important characteristic of these systems is that directed motion  is
realized by pure ac forces thanks to the convenient breaking of relevant symmetries.
\cite{Brouwer98,flach,lilipr}

\begin{figure}
\includegraphics[width=8cm,clip]{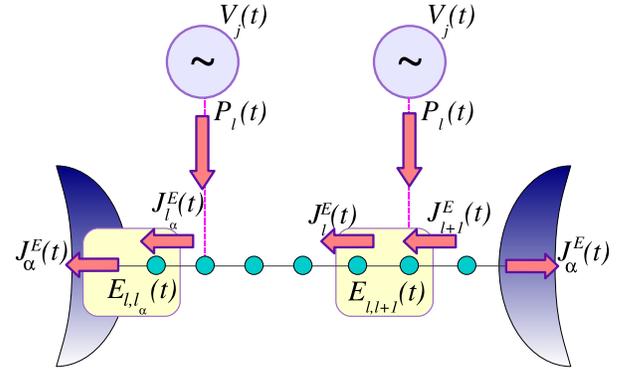}
\caption{(Color online) Sketch  of the device. The sites
  of the chain described by $H^{sys}(t)$ are represented by small circles.
 The local harmonically time-dependent fields
$V_j(t)$ develop a power $\overline{P}_j$. The right box
  encloses a unit of chain described by $H^{l,l+1}(t)$ where energy conservation is considered, while the
left one corresponds to a unit described by $H^{(cont)}$. The arrows
indicate the dc energy currents $\overline{J}^E_l$ flowing into a
bond $(l,l+1)$ of the system, $\overline{J}^E_{l_{\alpha}}$ entering
the contacts and $\overline{J}^E_{\alpha}$ entering the reservoirs.
} \label{fig1}
\end{figure}

Energy transport in stationary conditions is achieved as a response to temperature
and/or chemical potential gradients.  The application of time dependent fields
can induce  net particle transport between
 reservoirs at the same chemical potentials, while it brings
 about by itself heating of the sample. Then, it is possible that
  ac forces  can also generate directed heat
transport between those reservoirs even if they are
at the same temperatures.
If so: What determines the direction of that heat flow? In general:
What is the detailed energy balance for such a system? Is it possible to
transport part of this energy to develop work?
The theoretical study of the underlying physical processes demands a full
quantum-mechanical treatment of the problem to evaluate the heat
currents through the different parts of the device, as well as the calculation
of the powers developed by the external fields.  The aim of this
work  is to present a theoretical approach based in non-equilibrium Green's function
that will allow us to address details of the energy balance
in the framework  of an exactly solvable  model
of an electronic quantum pump.\cite{lilip}

The device can be described in terms of a Hamiltonian with three pieces
representing the electronic driven system, the contact to the reservoirs
and the reservoirs: $H(t)
= H^{sys}(t) + H^{cont} +  H^{res}$. For simplicity, we consider a two terminal
setup with left ($L$) and right  ($R$) reservoirs, and a $N$-site  one-dimensional lattice model
for the driven system. We assume that the latter has
  an  energy profile $\varepsilon^0_l$
and nearest neighbor hopping elements
$w_l$. At $M$ lattice  positions, ac potentials of the form
$V_j(t)= V_j \cos( \Omega_0 t + \varphi_j)$ are locally applied.
This system Hamiltonian can, thus, be expressed as
 $ H^{sys}(t) =  \sum_{l=1}^{N-1} H^{(l,l+1)}(t)$, being
\begin{eqnarray}
 H^{(l,l+1)}(t)&=& \varepsilon_l(t) c^{\dagger}_{l} c_{l}+\varepsilon_{l+1}(t)
 c^{\dagger}_{l+1} c_{l+1}\nonumber \\
& &-w_{l} \Big( c^{\dagger}_{l}c_{l+1}+ c^{\dagger}_{l+1}c_{l} \Big),
\label{hsys}
\end{eqnarray}
with $\varepsilon_{l}(t)= \zeta_l \Big( \varepsilon^0_l+\sum_{j=1}^M
\delta_{l,l_j} V_j(t) \Big) $, being $\zeta_l=1/2$ for $1<l<N$ and $\zeta_{l_{\alpha}}=1$,
 for the sites
that intervene in the connection to the reservoirs
 $l_{\alpha}\equiv 1,N$.
 The contacts are represented
by hopping terms between the reservoirs and the latter positions:
$H^{cont}=  \sum_{ \alpha, k_{\alpha}  } w_{\alpha}
(c^{\dagger}_{k_{\alpha}} c_{l_{\alpha}} + H.c.)$.
The reservoirs are described by free-electron models:
$H^{res}= \sum_{k_{\alpha}} \varepsilon_{k_{\alpha}}
c^{\dagger}_{k_{\alpha}} c_{k_{\alpha}}$, which are assumed to
remain at equilibrium with well defined chemical potentials $\mu_{\alpha}$ and
temperatures $T_{\alpha}$, even after being connected to the driven structure.

In order to define currents of energy, we can follow the quantum mechanical
counterpart of
the procedure carried out in Refs. \cite{cam,pere} for a classical
model. We
formulate the equation for the
conservation of the energy $ E_{l,l+1}(t)=\langle H^{(l,l+1)}(t) \rangle$
stored in a bond $(l,l+1)$.
The variation in time of this mean value is calculated by recourse to
Ehrenfest's theorem which casts:
\begin{eqnarray}
& & \frac{d  E_{l,l+1}(t)}{dt}=J^{E}_{l+1}(t)-J^{E}_{l}(t)  + 
\frac{1}{2} \Big( P_{l}(t)+P_{l+1}(t) \Big) ,  \label{contl}\\
& & \frac{d E_{\alpha,l_{\alpha}}(t)}{dt}= J^{E}_{l_{\alpha}}(t)-
J^{E}_{\alpha}(t)+ \frac{1}{2} P_{l_{\alpha}}(t), \label{contal}
\end{eqnarray}
where the first equation corresponds to bonds of
the system without sites in contact to reservoirs (i.e. $1<l<N-1$),
as the one enclosed by the right box of  Fig. \ref{fig1},
while the second one corresponds
 to the left bond that establishes the contact between system and reservoir
(see left box of Fig. \ref{fig1}).
The first two terms of (\ref{contl}) represent a discretized version of the divergence of
the energy current flowing through the bond, while the
last two terms represent the power developed by the
external forces and are equal to  $\langle \partial
  H^{(l,l+1)}(t)/\partial t \rangle$. In Eq. (\ref{contal}), the
first current represents the flow of energy towards the system, the
 second one is the flow of energy entering the reservoir, and the third one, the
power developed by the
external forces.

Denoting $\rho_{j,j'}(t)=\langle c^{\dagger}_{j'}(t)  c_j(t ) \rangle $,
the explicit expressions for the different energy
currents read:
\begin{eqnarray}
J^E_{l}(t) &=& 2 \mbox{Im}\Big\{\varepsilon_{l}(t)
\Big[  \rho_{l,l-1}(t)w_{l-1} - \rho_{l,l+1}(t)w_{l}  \Big] \label{curl}\\
& &
+w_{l-1}\rho_{l-1,l+1}(t) w_{l}\Big\},
\nonumber  \\
J^E_{{\alpha}}(t) &=&
-2 \mbox{Im}\Big\{
\sum_{k_{\alpha}}w_{\alpha}
\rho_{l_{\alpha},k_{\alpha}}(t) \varepsilon_{k_{\alpha}} \Big\}, \label{cural}
\end{eqnarray}
and a similar expression for $J^E_{l_{\alpha}}(t)$,
while the power developed by the external forces is:
\begin{equation}\label{power}
{P}_{l}(t)=  \sum_{j=1}^M  \delta_{l,l_j} \frac{dV_j(t)}{dt} \rho_{l_j,l_j}(t).
\end{equation}
In order to analyze the energy balance we focus on the dc components of the energy
currents and powers done by the external forces. The conservation of the energy
(\ref{contl}) and (\ref{contal}) implies
$\overline{dE_{l,l+1}(t)/dt}=0$ and $ \overline{d E_{\alpha,l_{\alpha}}(t)/dt}=0$,
which defines continuity equations for the dc energy currents
$\overline{J}^E_{l}$, $\overline{J}^E_{\alpha}$
and powers $\overline{P}_{l}$,
where $\overline{A} \equiv 1/\tau_0 \int_0^{\tau_0} dt
A(t)$, being $\tau_0=2 \pi/\Omega_0$
 the period of the oscillating
time-dependent fields.

To evaluate the different energy currents  and powers we employ the treatment based in Keldysh
non-equilibrium  Green's functions of Ref. \cite{lilip}, which
 has been useful to study charge
transport in the kind of systems we are considering.
The mean values of observables  entering the corresponding expressions
can be expressed in terms of lesser Green's functions as follows
$\rho_{j,j'}(t)=-i G^<_{j,j'}(t,t)$. The latter is evaluated from a Dyson
equation, which
for $j, j'$ lying on the central system reads:
\begin{eqnarray}\label{lesl}
 G^{<}_{l,l'}(t,t) &=& \sum_{\alpha=L,R} \sum_{kk'=-\infty}^{\infty} e^{-i k \Omega_0 t}
\int_{-\infty}^{\infty} \frac{ d \omega}{2 \pi} \Gamma^<_{\alpha}(\omega)
\nonumber \\
& &
\times  {\cal G}_{l,l_{\alpha}}(k+k',
\omega) {\cal G}^*_{l',l_{\alpha}}(k', \omega),
\end{eqnarray}
being $ \Gamma^<_{\alpha}(\omega)= i f_{\alpha}(\omega)
\Gamma_{\alpha}(\omega)$,  where   the Fermi function $
f_{\alpha}(\omega)$ depends on  $\mu_{\alpha}$ and $T_{\alpha}$, and
$\Gamma_{\alpha}(\omega)=2 \pi \sum_{k_{\alpha}}
|\epsilon_{k_{\alpha} l_{\alpha} }|^2 \delta(\omega
-\varepsilon_{k_{\alpha}})$. The Green's function ${\cal
G}_{l,l'}(k,\omega)$ is the $k$-th Fourier coefficient of the
Fourier transform of the retarded Green's function, which can be
exactly evaluated with
convenient methods. \cite{lilip}
 The above expression (\ref{lesl})
can be used in (\ref{curl}) and (\ref{power}) to evaluate
$\overline{J}^E_{l}$ and $\overline{ P}_{l_j}$.
Using properties of the Green's function \cite{lilimich}
 the dc component of the energy current flowing into the reservoir reads:
\begin{eqnarray}
 \overline{J}^E_{\alpha}
&=&\sum_{\beta=L,R} \sum_{k=-\infty}^{\infty}\int \frac{d \omega}{2 \pi}
(\omega +  k \Omega_0) \Gamma_{\alpha}(\omega +  k \Omega_0) \Gamma_{\beta}(\omega )
 \nonumber \\
& &
\times |{\cal G}_{l_{\alpha},l_{\beta}}(k,\omega)|^2
\Big[ f_{\beta}(\omega )-f_{\alpha}(\omega +  k \Omega_0) \Big].
\label{heatcon}
\end{eqnarray}

We are interested in the case of reservoirs at the same chemical potentials
$\mu_{\alpha}=\mu, \forall \alpha$.
Following the same line as in stationary transport at linear response,\cite{bee} we define
the dc-heat current as
$ \overline{J}_{\gamma}^Q= \overline{J}_{\gamma}^E -\mu
\overline{J}_{\gamma}^C$,
 the latter term being the convective flow which
depends on the dc-charge current $\overline{J}_{\gamma}^C$.\cite{note}
As this current is conserved, $\overline{J}_{\gamma}^Q$
obeys the same continuity equation satisfied by
$\overline{J}_{\gamma}^E$.
\begin{figure}
\includegraphics[width=9cm,clip]{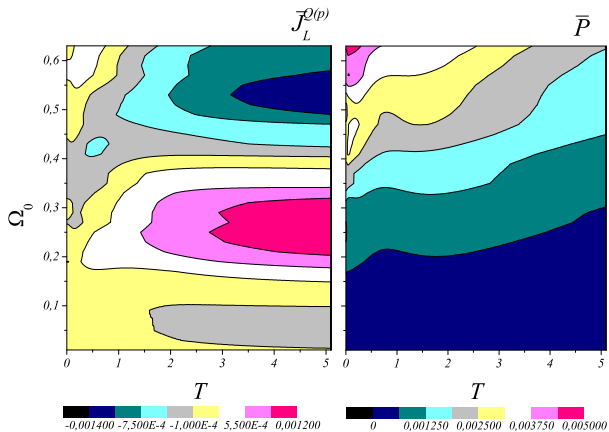}
\caption{(Color online) Pumped heat current (left) and  total power
$P= \overline{P}_1+ \overline{P}_2$
  developed by the forces (right) as functions of temperature and the pumping
  frequency.
Parameters are $N=22$, $E=1$, $V=0.2$, $T=0$, $\varphi=\pi/2$ and $\mu=-1.5$.}
\label{fig2}
\end{figure}
We now use the above theoretical framework  to analyze  two generic
 mechanisms that can take place in quantum pumps.
 The first one concerns the possibility of achieving {\em directed heat
 transport} between
reservoirs. To this end, notice,
 that the dc-heat
current along the lead $\alpha$ can be splitted as the addition of a generated
and a pumped contribution: $ \overline{J}_{\alpha}^Q= \overline{J}_{\alpha}^{Q (g)}+
\overline{J}_{\alpha}^{Q (p)}$, where the pumped component reads explicitly:
\begin{eqnarray}
 \overline{J}^{Q (p)}_{\alpha}
& = &\int \frac{d \omega}{2 \pi} (\omega -\mu ) \Big\{
 2  \Gamma^<_{\alpha}(\omega ) \mbox{Im}[{\cal
  G}_{l_{\alpha},l_{\alpha}}(0,\omega)] + \\
& &
\sum_{\beta=L,R} \sum_{k=-\infty}^{\infty}
 \Gamma_{\alpha}(\omega +  k \Omega_0) \Gamma^<_{\beta}(\omega )
|{\cal G}_{l_{\alpha},l_{\beta}}(k,\omega)|^2
  \Big\}.  \nonumber
\label{pum}
\end{eqnarray}
This component can be
proved to satisfy $ \sum_{\alpha=L,R} \overline{J}^{Q (p)}_{\alpha}=0$,
meaning that heat can be extracted from a given reservoir and injected into the other one.
From the continuity equation for the dc heat current, it follows that
$\sum_{j=1}^M \overline{ P}_{l_j}=\sum_{\alpha=L,R} \overline{J}^{Q(g)}_{\alpha}$,
indicating that the $\overline{J}^{Q(g)}_{\alpha}$ contribution accounts for the
heat generated by the external forces, which is dissipated into the reservoirs.

An example of the behavior of these two different contributions to the heat
flows at the reservoirs is shown in Fig. \ref{fig2} for a two barrier setup
in contact to reservoirs at the same temperature $T_L=T_R=T$.
We consider $\varepsilon^0_{l_j}=E$, $l_j=2, N-1$ and $\varepsilon^0_{l_j}=0, l \neq l_j$,
with two oscillating
potentials with the same amplitude  $V$ and
a phase-lag $\varphi_1=\varphi$,$\varphi_2=0$ applied at the barriers.
 Under such conditions a dc charge current $\overline{J}^C$ is induced
which behaves like $\overline{J}^C \propto V^2 \sin(\varphi)$ at small $V$.
\cite{Brouwer98}
The pumped component of the heat current, shown in the left panel of Fig. \ref{fig2},
 flows outwards in a reservoir and
inwards in the other one.
 It exhibits a complex structure of maxima, minima and sign
inversions as a function of $\Omega_0$. The details of these features are model-dependent.
They are consequence of the electronic propagation  through a
structure with discrete energy levels and quantum interference that takes place when
the pumping frequency $\Omega_0$ is resonant with the energy difference between these
levels. \cite{mobu}  The generated heat currents at the different leads
$\overline{J}^{Q  (g)}_{L,R}$,  also display a complex landscape in the
$T, \Omega_0$  plane. Their sum is, however, always positive and equals the
total developed power $\overline{P}=\overline{P}_1+ \overline{P}_2$, which
is a monotonous increasing function of $\Omega_0$ and decreases with
$T$ as shown in the right panel of Fig. \ref{fig2}.
\begin{figure}
\includegraphics[width=9cm,clip]{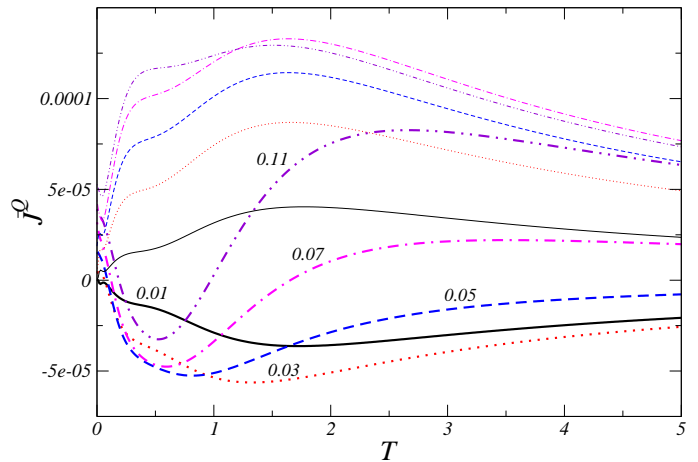}
\caption{(Color online) Total heat current as functions of $T$
for different pumping frequencies. Thin and thick lines
correspond to $\overline{J}^Q_L$ and  $\overline{J}^Q_R$, respectively.
Parameters are as in Fig. \ref{fig2}. The values of $\Omega_0$ are
indicated
in the plots.}
\label{fig3}
\end{figure}
At $T=0$ dissipation dominates and masks the heat pumping effect. For our example,
the
total heat currents
$\overline{J}_L^Q, \overline{J}_R^Q$ as functions of $T$ are plotted in Fig. \ref{fig3}.
There it is seen that the sign of both currents is positive at $T=0$,
which indicates that the flow goes from the central
system towards the reservoirs. Such a behavior is the one expected from
 considerations based on general
thermodynamics since, at $T=0$ there is no heat at the reservoirs amenable to
be transported.
 However, the results of Fig. \ref{fig3} show
that at  finite $T$ a regime exists for low pumping frequencies
$\Omega_0$ where heat pumping takes place. In fact,
the signs of the heat currents are different, indicating that they
leave one of the reservoirs and enter the other one. For higher $\Omega_0$
dissipation is again dominant and heat flows  always from the central
system into the two reservoirs.

\begin{figure}
\includegraphics[width=8cm,clip]{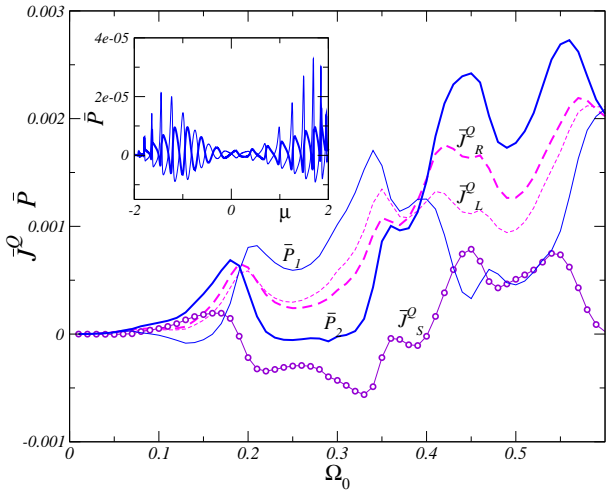}
\caption{(Color online) Main Panel: Heat currents $\overline{J}^Q_L, \overline{J}^Q_R$ (dashed thin
 and thick lines) and $\overline{J}^Q_S$ (circles) along
 the left and right leads and the central system, respectively for$\mu=-1.5$.
Powers developed by the external fields $\overline{P}_1, \overline{P}_2  $ (thin and
 thick solid lines). Inset:  $\overline{P}_1, \overline{P}_2  $
as function  of $\mu$ for $\Omega_0=0.01$.
 Parameters are as in Fig. \ref{fig2}.}
\label{fig4}
\end{figure}
The second remarkable mechanism we would like to analyze is the possibility of extracting work from the system.
This issue can be addressed on the basis of a perturbative solution of the Dyson equation
which allows for the evaluation of $\overline{P}_{l_j} $
at the lowest order in $V_j$.\cite{lilip,lilipr} It casts:
\begin{equation}\label{pert}
\overline{P}_{l_j} \sim \sum_{i=1}^M \Big[ \lambda^{(1)}_{ji}
\cos(\varphi_j- \varphi_i) +  \lambda^{(2)}_{ji}
\sin(\varphi_j- \varphi_i) \Big],
\end{equation}
where the coefficients
are
\begin{eqnarray}
\lambda^{(1)}_{ji}& = & \Omega_0 \frac{  V_i V_j}{2} \int_{-\infty}^{+\infty}
\frac{d  \omega}{2 \pi}
\mbox{Im} \Big\{ \gamma_{ji}(\omega) \gamma^{-}_{ji}(\omega)
 \Big\},\nonumber \\
\lambda^{(2)}_{ji}& = & \Omega_0 \frac{ V_i V_j}{2} \int_{-\infty}^{+\infty}
\frac{d  \omega}{2 \pi}
\mbox{Re} \Big\{ \gamma_{ji}(\omega)  \gamma^{+}_{ji}(\omega)
 \Big\},\nonumber \\
\end{eqnarray}
with $\gamma_{ji}(\omega)=\sum_{\alpha=L,R} \Gamma^<_{\alpha}( \omega) [G^0_{l_j,l_{\alpha}}(\omega)]^*
G^0_{l_i,l_{\alpha}}(\omega)$ and
$\gamma^{\pm}_{ji}(\omega)= \Big[G^0_{l_j,l_i}(\omega+\Omega_0)
\pm G^0_{l_j,l_i}(\omega-\Omega_0) \Big]$, being
 $G^0_{l,l'}(\omega)$ the equilibrium retarded
Green's function of the system in contact to the reservoirs but without the
time-dependent fields.
In the low frequency limit, $\lambda^{(1)}_{ij} \propto
\Omega_0^2$ while  $\lambda^{(2)}_{ij}\propto \Omega_0$ .
Thus, this solution indicates that quantum coherence in the
wave function propagation along the structure, which rules the behavior of
 the charge current, also plays a role in the way in which
energy is provided and exchanged. In particular, for more than one oscillating fields,
the terms $\propto \lambda^{(2)}_{ij}$ dominate at low enough $\Omega_0$. Since these terms
can be positive for some fields and negative for other ones, this
enables
a scenario where the total energy $\overline{P}=\sum_{j=1}^M \overline{P}_j$
is dissipated to the reservoirs at
a ratio $\propto \Omega_0^2$, while a larger amount of energy   $\propto\Omega_0$ is exchanged
between the different pumping centers. Such an effect is, in fact, observed
for some parameters in the example of the
two-barrier setup. Results are shown
in Fig.  \ref{fig4} for $T=0$ with $\varphi=\pi/2$. The dc powers are shown
 along with the dc heat currents
$\overline{J}^Q_L, \overline{J}^Q_R$ flowing to
the reservoirs and the one flowing within the system between the two barriers $\overline{J}^Q_S$.
The exchange of energy between the two fields is further highlighted in the inset
where the two dc powers $\overline{P}_1$ and $\overline{P}_2$ as
functions of $\mu$ are shown.
It is also interesting to note that the direction
of the heat flow between the two pumping centers, goes from the field doing the
largest power towards the other one. Instead, in the reservoirs the heat
 currents are always
positive, indicating that they flow inwards. These features are in line with the idea
that the fields heat locally the sample inhomogeneously, then the heat current
flows from the hottest to the coldest regions.

To conclude, we introduced a treatment based on the local conservation
of the energy in order to investigate details of
the energy transport in open quantum systems driven by time-dependent fields.
We identified two interesting mechanisms like the possibility of
achieving directed transport of heat between reservoirs at finite
but identical temperature as well as  extracting useful energy to
make work against the external forces. We illustrated these effects
in an exactly solvable model of an electronic quantum pump. Our
results  for reservoirs at $T=0$ also suggest that the direction of
the heat flows seem to  be ruled by an effective heat gradient
induced by the application of the external driving.

We thank M. B\"uttiker for stimulating conversations. This work is
supported by  PICT 03-11609 and CONICET from Argentina, BFM2003-08532-C02-01
and the ``Ramon y Cajal'' program (LA) from MCEyC of Spain.


\begin{thebibliography}{99}
\bibitem{note} This procedure has also been followed in the context of
  time-periodic transport for the calculation of heat current flowing into the
  leads (see \onlinecite{mobu,rey}).
\end{thebibliography}
\end{document}